# Modelling and Design of a Microstrip Band-Pass Filter Using Space Mapping Techniques

Saeed Tavakoli, Mahdieh Zeinadini, Shahram Mohanna

**Abstract**—Determination of design parameters based on electromagnetic simulations of microwave circuits is an iterative and often time-consuming procedure. Space mapping is a powerful technique to optimize such complex models by efficiently substituting accurate but expensive electromagnetic models, fine models, with fast and approximate models, coarse models. In this paper, we apply two space mapping, an explicit space mapping as well as an implicit and response residual space mapping, techniques to a case study application, a microstrip band-pass filter. First, we model the case study application and optimize its design parameters, using explicit space mapping modelling approach. Then, we use implicit and response residual space mapping approach to optimize the filter's design parameters. Finally, the performance of each design methods is evaluated. It is shown that the use of above-mentioned techniques leads to achieving satisfactory design solutions with a minimum number of computationally expensive fine model evaluations.

## 1 Introduction

Considering the development of computer-aided design techniques, optimization plays a vital role in modelling and design of microwave circuits. A typical design problem is to choose the design parameters to get the desired response. Space mapping (SM) approach, introduced by Bandler et al. [1], is a powerful technique to optimize complex models. It substitutes efficiently expensive electromagnetic models, fine models, with fast and approximate models, coarse models. To obtain the optimal design for the fine model, the SM establishes a mapping between parameters of the two models iteratively [1, 2]. SM techniques can be classified to original or explicit SM [3] and implicit SM (ISM) [4] methods. Both methods use an iterative approach to update the mapping and predict new design parameters.

Explicit space mapping modelling approach is based on setting up a surrogate model. SM-based surrogate models involve only certain combinations of input and output mappings. The input mapping is an explicit mapping between design parameters of the coarse and fine models. It is aimed to match SM-based surrogate and fine models in a region of interest.As evaluation of fine models is expensive, surrogate models, which should be fast, accurate and valid in a wide range of parameters, should be constructed using only a few fine model evaluations. In other words, SM-based surrogate models should use a small amount of data from fine models to extract the input and output's mapping parameters. Having the space mapping parameters established, the evaluation of SM-based surrogate models is approximately done using that of coarse models. This approach permits the creation of library models that can be used for model enhancement of microwave elements [5].

In some cases, the mapping established between parameters of the coarse and fine models is not explicit and it is hidden in the coarse model. This issue is addressed by ISM. The drawback of this approach is that ISM technique may not necessarily converge to the optimal solution. This problem can be solved using the ISM along with the response residual space mapping (RRSM) [6]. First, the algorithm starts with the ISM to reach a solution close to the optimal one. Then, RRSM approach is used to reach a satisfactory solution.

In this paper, explicit SM modelling approach as well as ISM and RRSM techniques are applied to a parallel-coupled-line microstrip band-pass filter. Agilent ADS [7] and Ansoft HFSS [8] are employed to simulate coarse and fine models, respectively.

## 2 Explicit Space Mapping Modelling

The modelling procedure starts with optimization of the coarse model to obtain the reference point of the region of interest. According to star distribution, shown in Figure 1, an $n-$dimensional interval centered at the reference point is created. Then, the input and output's mapping parameters are calibrated such that multiple sets of responses of the SM-based surrogate model match those of the fine model, simultaneously. To check the validity of the resulting model, it is tested with some test points in the region of interest.

• *Tavakoli is with the Faculty of Electrical and Computer Eng., The University of Sistan and Baluchestan, Iran.*
• *M. Zeinadini is with the Faculty of Electrical and Computer Eng., The University of Sistan and Baluchestan, Iran*
• *S. Mohanna is with the Faculty of Electrical and Computer Eng., The University of Sistan and Baluchestan, Iran.*

If the test results are not satisfactory, more data should be provided. The SM-based surrogate model's response is generated in specific points over a range of frequencies because the size of output mapping matrices is fixed. To generate the SM-based surrogate model's response over all points of the frequency range, optimized output mapping matrices should be interpolated using linear frequency interpolation techniques [5]. Then, the resulting model, provided for any frequency sweep, is optimized to determine optimal design parameters satisfying design specifications.

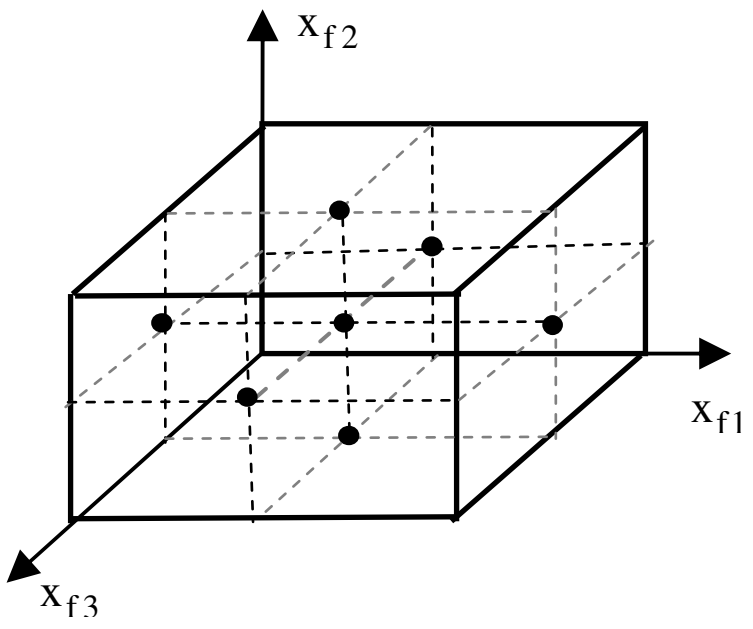

Fig. 1. Three-dimentional star set for the base points.

## 2.1 Model description

Let $R_f : X_f \rightarrow R^m$ and $R_c : X_c \rightarrow R^m$ refer to the fine and coarse model responses, respectively, where $X_f \subseteq R^n$ and $X_c \subseteq R^n$ are design parameter domains. For example, $R_f$ and $R_c$ may represent the magnitude of a transfer function of a microwave structure at $m$ chosen frequencies. Suppose that $X_R \subseteq X_f$ is the region of interest, in which we intend to enhance matching between the SM-based surrogate and fine models. As shown in Equation (1), $X_R$ is an $n-$dimensional interval in $R^n$ centered at reference point $x^0 \in R^n$.

$$X_R = [x^0 - \delta, x^0 + \delta] \quad (1)$$

where $\delta = [\delta_1 ... \delta_n]^T \in R^n$ determines the size of $X_R$. To obtain mapping parameters for which SM-based surrogate and fine models' responses become close enough, we use the star distribution. In this case, the set of evaluation points, known as the base set, consists of $(2n+1)$ points, where $n$ is number of design parameters. If the responses are not yet close enough, $2^n$ corner points may be added to the base set. Figure 2 shows the configuration of the SM-based surrogate and fine models.

The SM-based surrogate models can be given by

$$R_s(x, A, B, c, d) = A.R_c(B.x + c) + d \quad (2)$$

where $x$, and $R_s$ refer to design parameters of the surogate model and surrogate model response, respectively. $B$ and $c$ account for the input mapping [9], whereas $A$ and $d$ provide the output mapping [10]. Using the parameter extraction procedure, shown in Equation (3), these matrices are determined so that multiple sets of responses of the SM-based surrogate model simultaneously match those of the fine model.

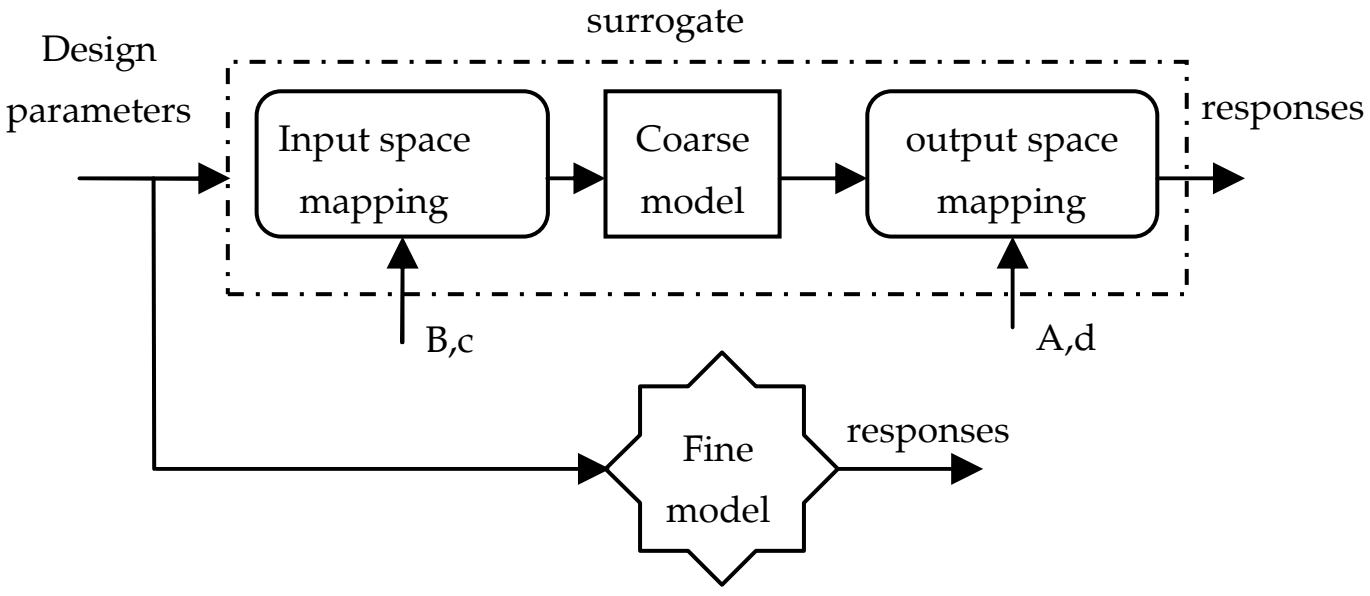

Fig. 2. Illustration of the SM-based model.

$$\left(\overline{A}, \overline{B}, \overline{c}, \overline{d}\right) = \arg \min_{(A,B,c,d)} \sum_{K=0}^{2n} \left\| R_f\left(x^K\right) - R_s\left(x^K, A, B, c, d\right) \right\| \quad (3)$$

To check the validity of the resulting SM-based surrogate model, then, it is tested using multiple test points [5].

## 3 Implicit and Response Residual Space Mapping

This approach starts with the ISM to reach a solution close to the optimal one. Then, RRSM approach is used to reach a satisfactory solution. First, the coarse model is optimized to obtain design parameters satisfying the design objectives. Second, an auxiliary set of parameters in the coarse model, which always remain fixed in the fine model, is calibrated to match the coarse and fine models. Examples of the auxiliary parameters are physical parameters such as relative dielectric constant and geometrical parameters such as substrate height. The coarse model with updated values of auxiliary parameters is known as the calibrated coarse, surrogate, model. Considering the re-calibrated auxiliary parameters fixed, then, the calibrated coarse model is re-optimized to obtain a new set of design parameters. These design parameters are given to the fine model to evaluate its performance [4, 11]. If responses of the resulting calibrated corase and fine models are not yet close enough, a new surrogate model is generated. It is created from the calibrated coarse model and a weighted residual term. This term is calculated using weighted misalignment between the fine and previous coarse models' responses. The previous coarse model is the calibrated coarse model in which the auxiliary and design parameters are fixed. Next, the new surrogate is re-optimized to determine a new set of design parameters. Finally, the resulting new set of design parameters is given to the fine model to evaluate its performance. The procedure is stopped when the fine model's response satisfies design specifications.

### 3.1 RRSM approach

The design objective is to calculate the following optimal solution for the fine model

$$x_f^* = \arg\min_{x_f} \Omega(R_f(x_f)) \quad (4)$$

where $\Omega$, $R_f$, $x_f$ and $x_f^*$ refer to the given objective function, fine model's response, fine model design parameters and optimal fine model design parameters, respectively. $x_f^*$ can be found using the following iterative procedure

$$x_f^{k+1} = \arg\min_{x_f} \Omega\left(R_c\left(x_f, p^k\right)\right) \quad (5)$$

where $p$ and $R_c(x_f, p)$ refer to the auxiliary parameters and a response vector of the coarse model, respectively. Using Equation (6), the auxilary parameters at the $k^{th}$ iteration, $p^k$, are obtained so that the surrogate and fine model's responses become similar enough. $R_c\left(x_f, p^k\right)$ denotes the surrogate model's response with these auxilary parameters.

$$p^k = \arg\min_{p} \left\| R_f\left(x_f^k\right) - R_c\left(x_f^k, p\right) \right\| \quad (6)$$

The initial surrogate model is $R_c\left(x_f, p^0\right)$, where $p^0$ represents initial auxiliary parameters. In other words, the surogate model is the coarse model with updated values of the auxiliary parameters [11]. In certain cases, the explicit or implicit space mapping thechniques may not converge to the optimal solutions. Using the RRSM aproach, however, the same cases converge [6]. The RRSM addresses residual misalignment between the responses of the optimal coarse and optimal fine models. Using RRSM technique, the new surrogate model is defined as follows

$$\begin{aligned} R_s &= R_c\left(x_f, p\right) + \mathrm{diag}\{\lambda_1, \lambda_2, \ldots\ldots\ldots\ldots\lambda_m\}\Delta R, \\ \Delta R &= R_f\left(x_f\right) - R_c\left(x_f^{*k}, p\right) \end{aligned} \quad (7)$$

where $R_c(x_f, p)$, $x_f^{*k}$ and $m$ refer to the response of calibrated coarse model, optimal design parameter of fine model at the $k^{th}$ iteration, and number of sample points, respectively. The residual term, $\Delta R$, is the difference between the previous calibrated coarse and fine models' responses. The previous coarse model is the calibrated coarse model in which the auxiliary and design parameters are fixed. $\Delta R$ is weighted by a weighting parameter $\lambda_i$, with $0 \le \lambda_i \le 1$ [6].

Considering the re-calibrated auxiliary parameters fixed, then, the new surrogate is re-optimized to obtain a new set of design parameters, $x_f^{k+1}$, in the next step.

$$x_f^{k+1} = \arg\min_{x_f} \Omega\left(R_s\right) \quad (8)$$

If the fine model's response for this new set of design parameters satisfies the design specifications, the algorithm is stopped. Otherwise, it re-calculates the auxiliary parameters for the current design parameters [4, 6].

## 4 Illustrative Example: Parallel-Coupled-Line Microstrip Band-Pass Filter

The structure of the parallel-coupled-line microstrip band-pass filter is illustrated in Figure 3.

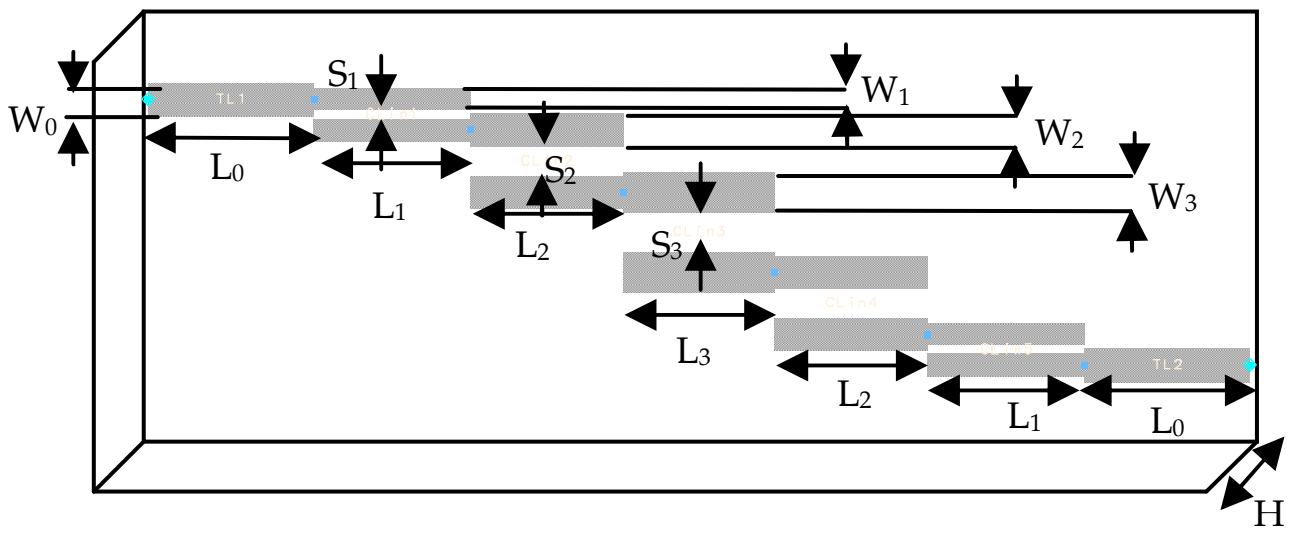

Fig. 3. Parallel-coupled-line microstrip band-pass filter.

The design specifications are as follows.

$|S11| \le -12\,\text{dB}, 8.9\ \text{GHz} \le \omega \le 10.1\ \text{GHz}$

$|S12| \le -30\,\text{dB}, \omega \le 8.2\ \text{GHz}\ \&\ \omega \ge 11.4\ \text{GHz}$

The coarse model is simulated by ADS as shown in Figure 4. As the filter has a symmetric structure, coupled lines CLin5 and CLin4 are identical to CLin1 and CLin2, respectively.

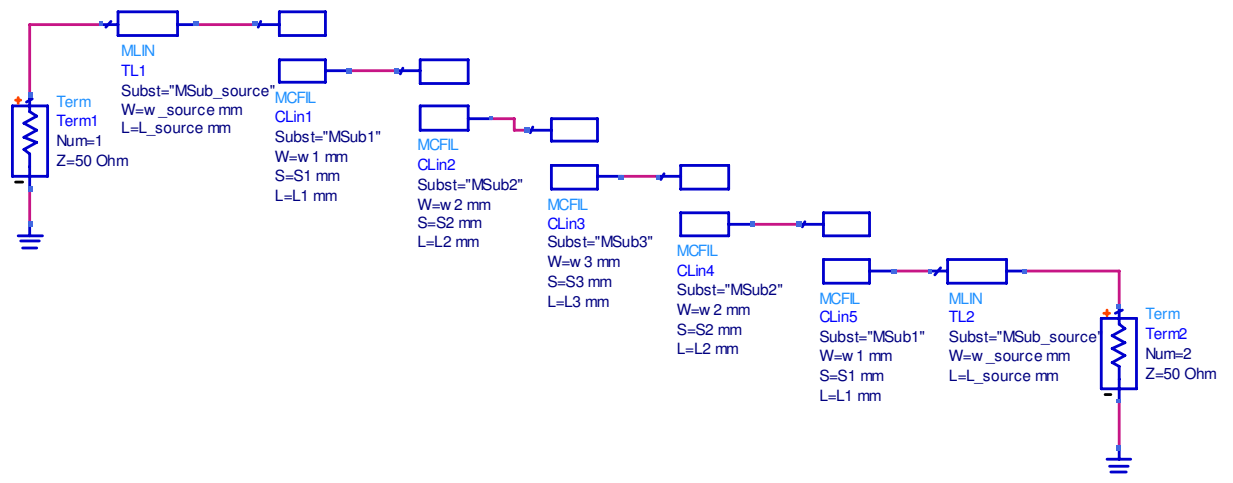

Fig. 4. Coarse model simulated by ADS.

The filter structure is made of a perfect conductor on the top of a substrate with a relative dielectric conestant of 10.2 and a height of 0.635 mm, backed with a perfect counductor ground plane. To simplify the modelling and design procedure by reducing the number of design parameters, the following parameters (all in mm) are assumed to be constant.

$w_0 = 0.59, w_1 = 0.383, w_2 = 0.575, w_3 = 0.595, L_0 = 3$

### 4.1 Explicit space mapping modelling

$x_f = [S_1\, L_1\, S_2\, L_2\, S_3\, L_3]^T$ and $x_c = [S_{1c}\, L_{1c}\, S_{2c}\, L_{2c}\, S_{3c}\, L_{3c}]^T$ refer to fine and coarse model parmeters, respectively. The region of interest is specified by the following regions

$0.1288 \le S_1 \le 0.1932$, $2.7661 \le L_1 \le 2.9372$,

$0.4320 \le S_2 \le 0.6480$, $2.6904 \le L_2 \le 2.8569$,

$0.5840 \le S_3 \le 0.8759$, $2.6751 \le L_3 \le 2.8406$

all in mm. The frequency range used is 8 to 12 GHz with a step of 0.25 GHz (17 points). The coefficient, $S_{12}$, is used to match responses of the SM-based and fine models. Since we have six independent design parameters, the number of base set points in the region of interst is 13.

The reference point obtained from coarse model optimization is given by

$$x^0 = [0.161, 2.8517, 0.54, 2.7737, 0.73, 2.7579]^T$$

The linear input mapping is given by

$$S_{1c} = b_{11}S_1 + b_{12}L_1 + b_{13}S_2 + b_{14}L_2 + b_{15}S_3 + b_{16}L_3 + c_1$$
$$L_{1c} = b_{21}S_1 + b_{22}L_1 + b_{23}S_2 + b_{24}L_2 + b_{25}S_3 + b_{26}L_3 + c_2$$
$$S_{2c} = b_{31}S_1 + b_{32}L_1 + b_{33}S_2 + b_{34}L_2 + b_{35}S_3 + b_{36}L_3 + c_3$$
$$L_{2c} = b_{41}S_1 + b_{42}L_1 + b_{43}S_2 + b_{44}L_2 + b_{45}S_3 + b_{46}L_3 + c_4$$
$$S_{3c} = b_{51}S_1 + b_{52}L_1 + b_{53}S_2 + b_{54}L_2 + b_{55}S_3 + b_{56}L_3 + c_5$$
$$L_{3c} = b_{61}S_1 + b_{62}L_1 + b_{63}S_2 + b_{64}L_2 + b_{65}S_3 + b_{66}L_3 + c_6$$

The initial input and output mapping matrices are given by $B_{6\times6} = I$, $c_{6\times1} = 0$ and $A_{17\times1} = 1$, $d_{17\times1} = 0$, respectively. Quasi-Newton optimization algorithm is used to model the filter and obtain final mapping parameters, as follows.

$$A^T = \begin{bmatrix} 0.7916, 0.8606, 0.9473, 0.8541, 0.8227, \ldots \\ 0.9201, 0.9696, 0.9828, 0.9795, 0.9824, \ldots \\ 1.0081, 1.0442, 1.0121, 0.9502, 0.8955, \ldots \\ 0.8514, 0.8182 \end{bmatrix}$$

$$B = \begin{bmatrix} 95.73 & 0.25 & 0.76 & -0.45 & 1.92 & -2.87 \\ -0.51 & 97.75 & 3.297 & -0.93 & 0.53 & -2.25 \\ 0.83 & -1.10 & 87.01 & -0.57 & -1.83 & 0.83 \\ -0.009 & 0.77 & 16.55 & 94.64 & 3.81 & -2.92 \\ 0.80 & -1.05 & -1.75 & -0.29 & 88.98 & 1.14 \\ 0.54 & 1.57 & 14.31 & -1.75 & 3.21 & 97.26 \end{bmatrix} \times 10^{-2}$$

$$c^T = [56, 1407, -0.2876, 7589, 1117, 3707] \times 10^{-4}$$

$$d^T = \begin{bmatrix} 79.38, 7.549, 91.81, 656.2, 87.59, \ldots \\ 3.219, 13.77, 107.8, 157.9, 83.93, \ldots \\ 27.42, 1803.7, 1686.4, 325.2, 90.59, \ldots \\ 37.52, 8.1044 \end{bmatrix} \times 10^{-15}$$

Error plots for the coarse and SM-based models in some test points are shown in Figures 5 and 6. For one test point, the magnitude of $S_{12}$ for the fine, coarse, and surrogate models is depicted in Figure 7. After creating SM-based model, it is optimized with respect to design paramers.

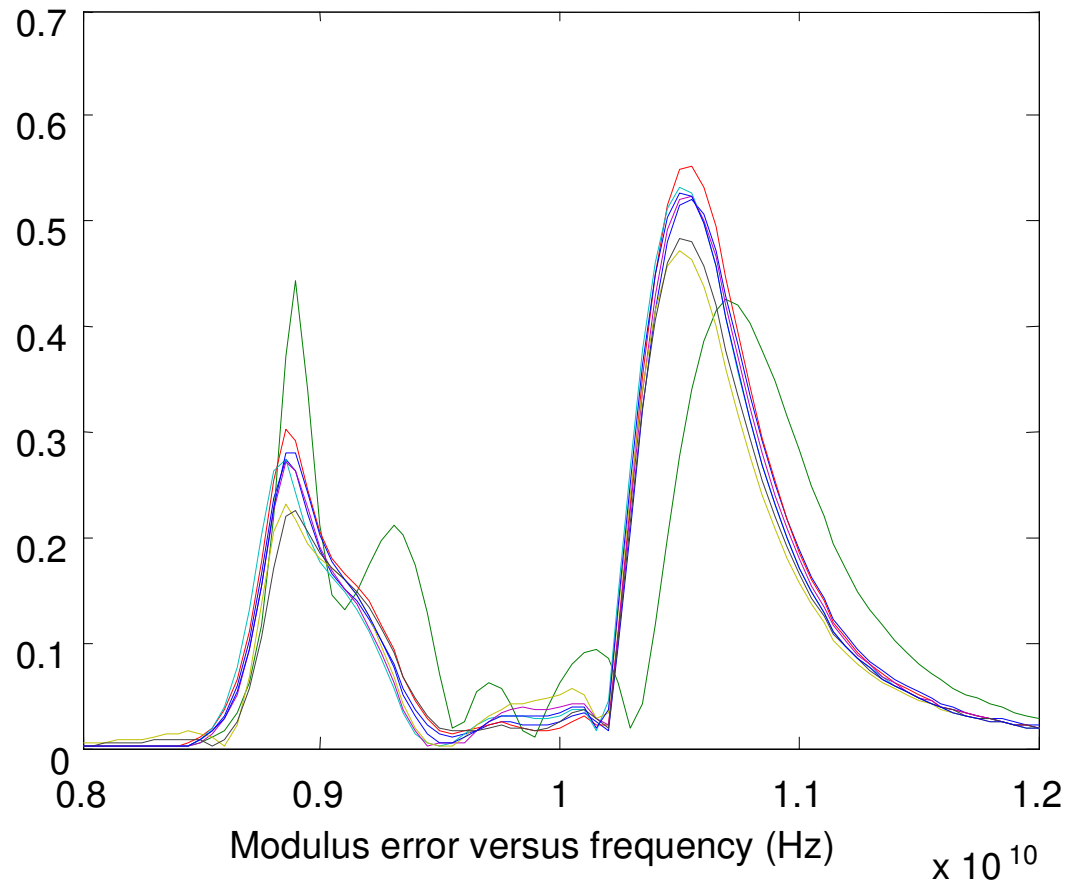

Fig. 5. ADS error plots for the coarse model (modulus of difference between $R_c$ and $R_f$).

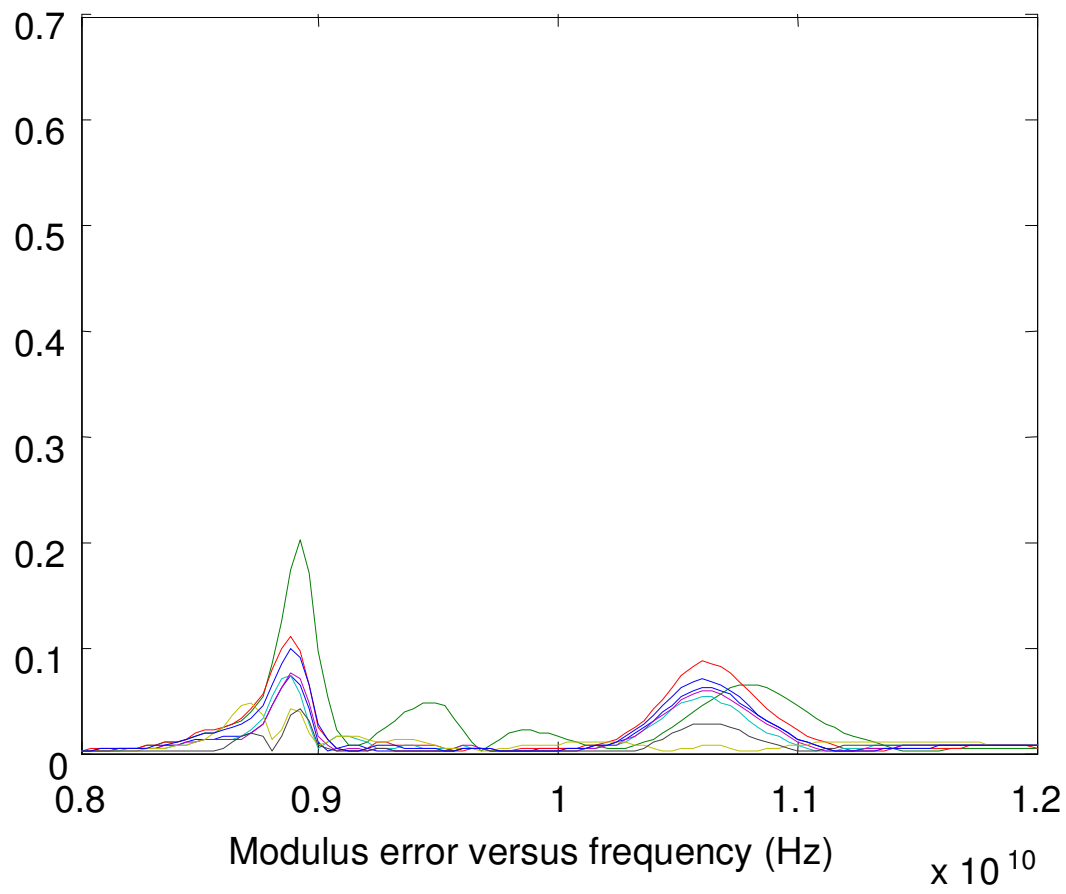

Fig. 6. ADS error plots for the surrogate model (modulus of difference between $R_s$ and $R_f$).

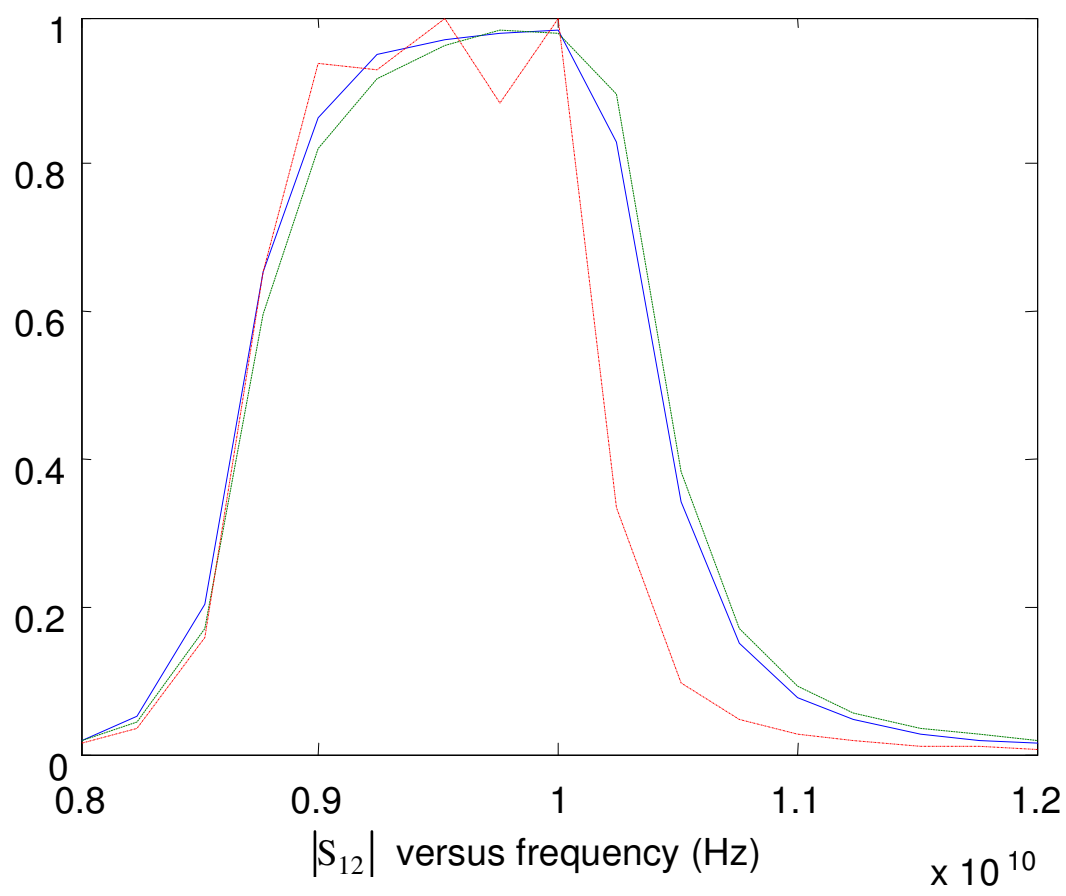

Fig. 7. $|S_{12}|$ for the fine (-), coarse (-.), and surrogate (--) models for one test point.

These parameres are, then, given to the fine model. For these parameters, Figure 8 shows the fine model response. Table 1 shows optimal design parameters, obtained from optimizing the SM-based model.

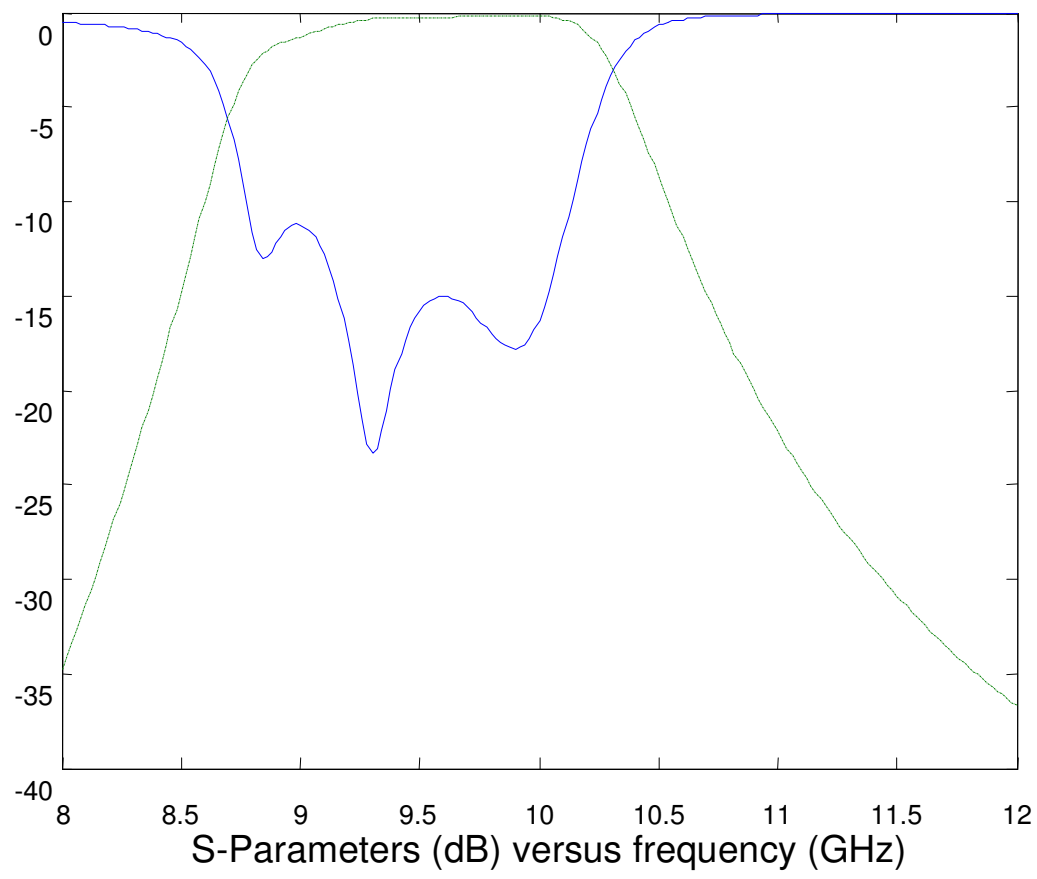

Fig. 8. Optimal fine model responses, $S_{11}$ (-), $S_{12}$ (-.), using HFSS.

TABLE 1
DESIGN PARAMETERS

| Design parameters | Optimal solution ( mm ) |
|---|---|
| $S_1$ | 0.19321 |
| $L_1$ | 2.93725 |
| $S_2$ | 0.63118 |
| $L_2$ | 2.85686 |
| $S_3$ | 0.79268 |
| $L_3$ | 2.71186 |

### 4.2 Implicit and response residual space mapping

The design parameters are the lengths of coupled lines and the distances between them, as follows.

$x_f = [S_1\, L_1\, S_2\, L_2\, S_3\, L_3]$

The auxiliary parameters are the heights of microstrip lines and their relative dielectric constants, as follows.

$p = [h_0\, h_1\, h_2\, h_3\, \varepsilon_{r0}\, \varepsilon_{r1}\, \varepsilon_{r2}\, \varepsilon_{r3}]$

where $h_i$ and $\varepsilon_{ri}$ refer to the height of $i^{th}$ microstrip line and its relative dielectric constant, respectively.

The following design procedure should be accomplished to obtain optimal design parameters.

Step 1. Set up the coarse model in ADS.

Step 2. Optimize the coarse model with respect to design parameters using ADS' gradient optimization algorithm.

Step 3. Simulate the fine model in HFSS, considering the solution given by ADS.

Step 4. Evaluate the fine model's response. The design procedure is terminated if the fine model's response satisfies the design specifications.

Step 5. Import fine model's response to ADS. Using Quasi-Newton optimization algorithm, the real and imaginary parts of scatterring parameters are used to match the fine model's response and that of either the calibrated coarse model (when using ISM) or the new surrogate model (when using RRSM). In the new surrogate model, the residual term, $\Delta R$, is weighted by weighting parameters. They are chosen as $\lambda_i = 0.5, i = 1,.......m$ . The model matching procedure is done by calibrating auxiliary parameters.

Step 6. Re-optimize either the calibrated coarse model or the new surrogate model to obtain next fine model's design parameters.

Step 7. Update the fine model and go to step 4.

The coarse and fine models' responses for the initial solution are shown in Figure 9. The algorithm requires two iterations, i.e. three fine model simulations. The coarse and fine model responses for the final solution are shown in Figure 10. Table 2 shows the initial and final design parameters.

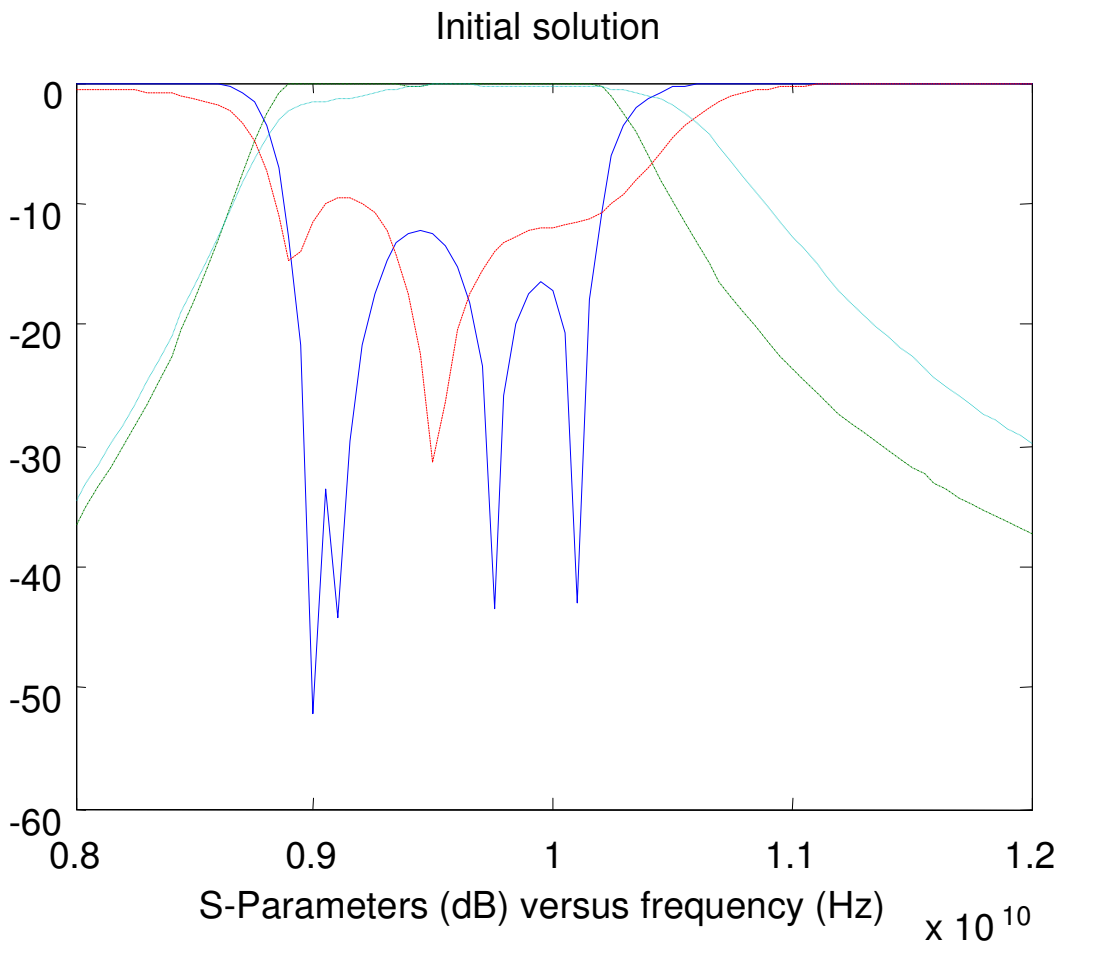

Fig. 9. $S_{11}$ -coarse (-), $S_{12}$ -coarse (-.), $S_{11}$ -fine (--), $S_{12}$ -fine (.).

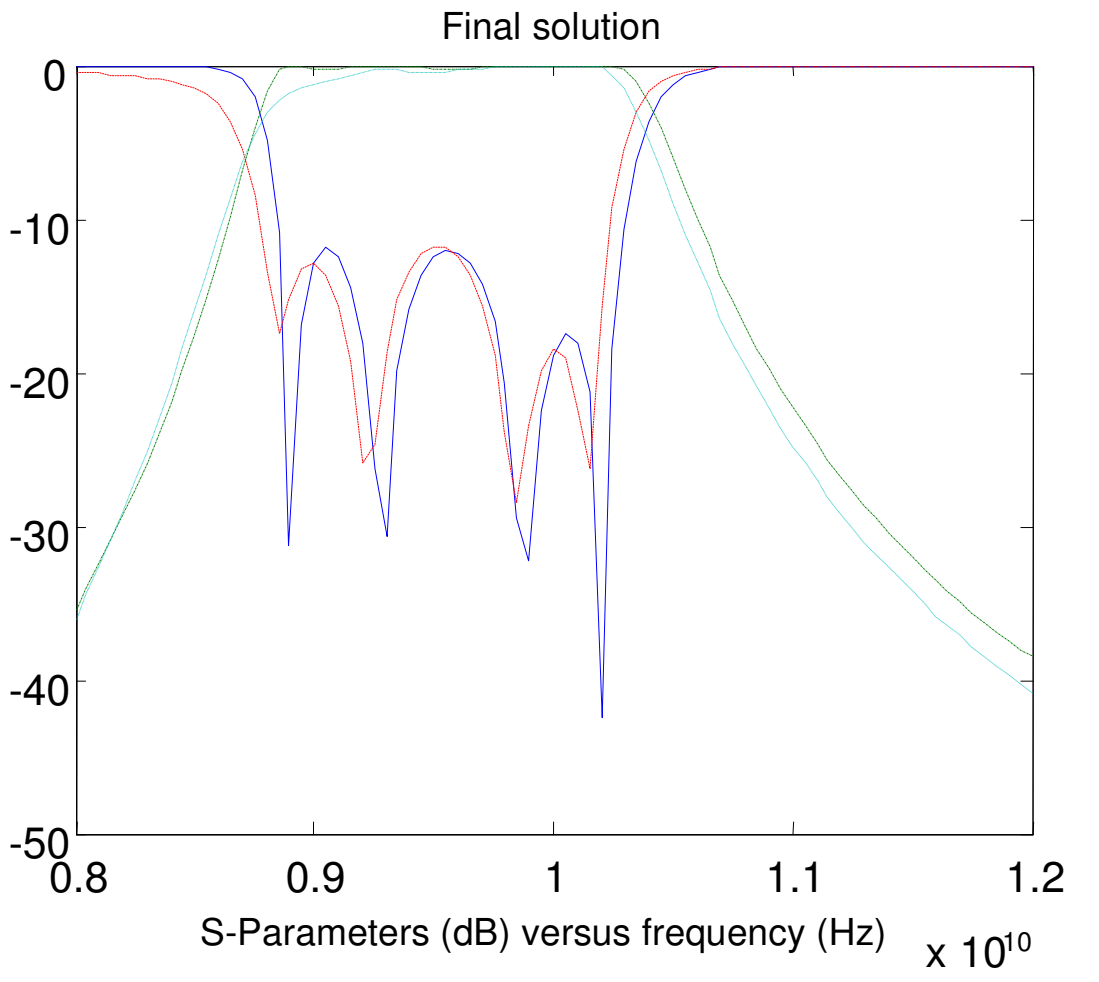

Fig. 10. $S_{11}$ -coarse (-), $S_{12}$ -coarse (-.), $S_{11}$ -fine (--), $S_{12}$ -fine (.).

TABLE 2
DESIGN PARAMETERS

| Design parameters | Initial solution ( mm ) | Final solution ( mm ) |
|---|---|---|
| $S_1$ | 0.16100 | 0.26492 |
| $L_1$ | 2.85170 | 3.33349 |
| $S_2$ | 0.53998 | 0.57697 |
| $L_2$ | 2.77365 | 2.43921 |
| $S_3$ | 0.72995 | 0.92947 |
| $L_3$ | 2.75786 | 2.98883 |

## 5 DISCUSSION

In SM-based modelling approach, the fine model should be simulated as much as $2n+1$ times. Therefore, the SM-based modelling is suitable for low-dimensional problems. One advantage of the ISM is that, unlike input space mapping, it does not affect the domain of the surrogate model, which may be important in the case of constrained optimization [4]. Also, since ISM method dose not require matrix calculations, it is probably the simplest technique to implement.

It can be seen from Figure 9 that when the design parameters obtained by ADS are given to HFSS, the bandwith is wider than necessary. It is not desirable as the increased bandwith can affect the pass-band performance. Using the implicit and response residual space mapping approach, Figure 10 shows that the final design parameters satisfy the design objectives and result in the requested bandwidth.

## 6 CONCLUSIONS

This paper aimed to model a parallel-coupled-line microstrip band-pass filter and optimize its design parameters. Using explicit space mapping modelling aproach, a surrogate model was used instead of fine one to simplify the design procedure. To get satisfactory results, the accuracy of coarse models should be good enough. First, a linear input mapping was used to match fine and coarse models' responses. As these responses were not close enough, an output mapping was also used. Simulation results demonstrated that the resulting SM-based surrogate model was fast and accurate enough.

Using ISM and RRSM approach, the procedure of design and optimization of the fine model using ADS was started with implicit space mapping. As the calibration step did not improve enough the match, a surrogate model was generated to establish an output mapping between mapped surrogate and coarse model responses. As a result, the responses of surrogate and fine models became close enough. Using this technique, the filter's design parameters were determined. This algorithm required one iteration of the ISM and one iteration of the RRSM. It was shown that only three evaluations of the fine model were sufficient to get satisfactory results.

**Saeed Tavakoli** received his BSc and MSc degrees in electrical engineering from Ferdowsi University of Mashhad, Iran in 1991 and 1995, respectively. In 1995, he joined the University of Sistan and Baluchestan, Iran. He earned his PhD degree in electrical engineering from the University of Sheffield, England in 2005. As an assistant professor at the University of Sistan and Baluchestan, his research interests are space mapping optimization, multi-objective optimization, control of time delay systems, PID control design, robust control, and jet engine control. Dr. Tavakoli has served as a reviewer for several journals including IEEE Transactions on Automatic Control, IEEE Transactions on Control Systems Technology, IET Control Theory & Applications, and a number of international conferences.

**Mahdieh Zeinadini** obtained her BSc degree in electrical engineering from Shahid Bahonar University of Kerman, Iran in 2007. Currently, she is an MSc student at the University of Sistan and Baluchestan, Iran. Her areas of research include space mapping optimization and design of microwave circuits.

**Shahram Mohanna** received his BSc and MSc degrees in electrical engineering from the University of Sistan and Baluchestan, Iran and the University of Shiraz, Iran in 1990 and 1994, respectively. He then joined the University of Sistan and Baluchestan, Iran. In 2005, he obtained his PhD degree in electrical engineering from the University of Manchester, England. As an assistant professor at the University of Sistan and Baluchestan, his areas of research include design of microwave circuits, antenna design and applied electromagnetic. Dr. Mohanna has served as a reviewer for several journals and a number of conferences.